# Attrition and Non-Response in Panel Data: The Case of the Canadian Survey of Labor and Income Dynamics

Brahim Boudarbat (University of Montreal)

Lee Grenon (Statistics Canada)

**Abstract**

This paper provides an analysis of the effects of attrition and non-response on employment and wages using the Canadian Survey of Labour and Income Dynamics. We consider a structural model composed of three freely correlated equations for non-attrition/response, employment and wages. The model is estimated using microdata from 22,990 individuals who provided sufficient information in the first wave of the 1996-2001 panel. The main findings of the paper are that attrition is not random. Attritors and non-respondents likely are less attached to employment and come from low-income population. The correlation between non-attrition and employment is positive and statistically significant, though small. Also, wage estimates are biased upwards. Observed wages are on average higher than wages that would be observed if all the individuals initially selected in the panel remained in the sample.

Corresponding author:

Brahim Boudarbat, Assistant Professor
School of Industrial Relations, Université de Montréal, C.P. 6128, Succursale Centre-ville, Montreal, QC, Canada H3C 3J7
Phone: +1-514-343-7320. Fax: +1-514-343-5764
Email: brahim.boudarbat@umontreal.ca

The authors are grateful for valuable comments and suggestions from Claude Montmarquette and participants in the 45[th] annual conference of the Canadian Society of Economics (Charlevoix, 2005) and Statistics Canada's 22[nd] International Methodology Symposium, Methodological Challenges for Future Information Needs (Ottawa, 2005)**.** The analysis presented here uses microdata from Statistics Canada. However, the interpretations and opinions expressed by the authors do not represent the opinions of Statistics Canada. Brahim Boudarbat also gratefully acknowledges financial support from SSHRCC INE Grant "Globalization, technological revolutions and education."

## 1. INTRODUCTION

The increased availability of longitudinal data surveys has significantly boosted the empirical studies aimed at analyzing the dynamics of individual behaviors. The combination of both cross-section and cross-time observations provides useful leverage for identifying the parameters that drive those behaviors, and allows controlling for individual-specific unobservable effects. In Canada, the longitudinal Survey of Labor and Income Dynamics (SLID) is increasingly used by researchers for the study of education, family, labor and income. The survey has the advantage of surveying very large samples of persons and providing detailed panel data (six annual waves) on a large variety of variables that are of interest for many research topics (see Section II). The value of the SLID and other sample surveys for empirical analyses is the reliable estimation of population statistics. The assumption of a representative sample is not valid if attrition and non-response over the course of the longitudinal panel are non-random.

This study examines the selectivity for attrition within model estimation for a specific subsample of longitudinal respondents. This study is not an assessment of non-response rates within SLID. Statistics Canada publishes information and measurement of data quality (Michaud and Webber, 1994; Armstrong and House, 2005) that are the best sources for assessment of non-response. In this study, we are concerned with longitudinal respondents who participated in the labor interview of panel 2 of the SLID (1996-2001) and were 16 to 64 years-old in the first year of the panel, i.e. in 1996. Among these working-age longitudinal respondents, 12.5% became out of scope during the panel (see Table 1). Respondents become out-of-scope when they migrate away from the Canadian provinces, are institutionalized, or are deceased. The out-of-scope respondents are outside of the target population for SLID, and are not eligible to participate for the reference year. The other component of attrition is non-response which includes persons who can not be located or contacted, and those who completely refuse to participate. The non-respondents are potentially still in the target population of the survey, but are no longer participating. In panel 2 of the SLID, complete non-response to the labor interview was at 9.4% at the end of the panel (2001) of working-age people who responded in the first wave (1996). The percentage who did not respond peaked at 11.4% in 2000. Thus, attrition and non-response behaviors concern over one-fifth (21.9%) of the working-age longitudinal sample who were in-scope and responded to the first wave of the labor interview (see Table 1).

As sampled individuals exit from the sample, the data set becomes less representative of the population from which the longitudinal sample was drawn if the attrition is non-random. Some econometric studies have analyzed the effect of attrition within longitudinal data on model estimation.[1] This literature generally indicates evidence that the labor market behavior of attritors and participants is different, although ignoring the selection bias has a minimal or negligible impact on estimation. Van Den Berg, Lindeboom and Ridder (1994) use panel data from The Netherlands and find that unobserved explanatory variables for the duration of panel survey participation of an individual are not related to unobserved explanatory variables for the duration of unemployment of that individual. In a subsequent study, Van Den Berg and Lindeboom (1998) find significant dependence between labor market durations and attrition, but there is little bias from ignoring this dependence. Lillard and Panis (1998) find, on the basis of data from the Panel Study of Income Dynamics (PSID), that despite the evidence of selectivity in attrition, the biases that are introduced by ignoring selective attrition are very mild. In the same vein, Zabel (1998)'s study which uses data from the PSID and the Survey of Income and Program Participation (SIPP), shows little indication of bias due to attrition in a model of labor market behavior, though there is evidence that the labor market behavior of attritors and participants is different.

---

[1] See for instance the Special Issue "Attrition in Longitudinal Surveys," of *The Journal of Human Resources*, Spring, 1998, Vol. 33, No. 2.

In this study, we aim at verifying whether the above consistent result regarding the effect of attrition on estimation applies to the SLID. For this purpose, we consider a structural model composed of three freely related equations for non-attrition/response, employment and wages. The relationship between these equations arises from the fact that the employment status is observed only for respondents to the labor interview, and that wages are observed only for the respondents who are employed. Thus, the model allows testing for the selectivity from attrition in both employment and wages equations. Our data is described in Section 2, the model is developed in Section 3, and structural parameter estimates are presented in Section 4. Our results provide evidence for non randomness of attrition/non-response behaviors. We find a positive and significant, though small, correlation between participating in the survey and being employed. Attritors and non-respondents are less likely to be employed, more "mobile" and have lower education levels. They also more likely to be immigrants and come from the lower end of the income distribution. With regards to the wages equation, we also find evidence for selection bias from both attrition and employment. Observed wages in the selected (available) sample likely are higher than wages in a sample drawn randomly with identical observed characteristics. We estimate the wages gap between the two samples at 9.65% due to attrition selectivity, and 13.51% due to employment selectivity. Section 5 offers a short summary with concluding remarks.

## 2. DATA

This study uses microdata from Statistics Canada's Survey of Labor and Income Dynamics (SLID). SLID is the principal household survey for information on income and one of the major data sources for labor dynamics of Canadian individuals and families. Historically, information on individual and family annual income was collected through the Survey of Consumer Finances (SCF), which was an annual cross-sectional survey of households in the ten provinces. Starting with the 1996 reference year, SLID replaced the SCF as the principal source of data on individual and family annual income, and the SCF was discontinued. SLID provides several important advantages over the SCF. First, SLID is a longitudinal survey that facilitates analysis of family, income and labor dynamics over time. Second, SLID collects the same income information as the SCF, however SLID adds a diverse range of information on transitions in jobs, income and family events.

SLID is a panel of longitudinal respondents selected at the beginning of the reference period and then interviewed annually for six years. The longitudinal sample is selected from households in the Labor Force Survey (LFS). The LFS is a multi-stage probability area sample. The longitudinal sample for SLID comprises all persons living in the selected households at the time of the LFS reference period. After the panel reference period begins, any cohabitant who lives with a longitudinal respondent for any period of the reference year is also interviewed. However, information for cohabitants is only collected for reference years in which they live with a longitudinal respondent. SLID permits proxy interviews and uses computer assisted telephone interviewing technology which enables feedback of information from previous interviews to reduce inconsistencies or "seam" problems between waves.

Each panel of the SLID has a relatively large sample size of approximately 15,000 households and 31,000 persons of 16 years of age or older, which is representative of the population of the ten provinces of Canada at the time of selection. Residents of the Yukon, Northwest Territories, and Nunavut, institutional residents, and persons living on Indian reserves are not eligible for selection into the longitudinal sample. Data collection for each reference year occurs in two phases. The first phase is a labor and household interview conducted shortly after the end of the reference year. The second data collection phase involves an income interview conducted later in the spring for the reference year. A substantial proportion of respondents agree to share their income tax information with Statistics Canada instead of responding to the income interview (Michaud and Latouche, 1996). SLID imputes values for selected income variables for non-response when sufficient information is available for other characteristics of the respondent (Webber and Cotton, 1998).[2]

Analyses with SLID will typically use all selected respondents who are in scope at the end of the selected reference period. Respondents who are out of scope at the end of the selected reference period are typically excluded from analyses and have a final longitudinal weight equal to zero. Longitudinal respondents may become out of scope due to institutionalization, migration out of the ten provinces, or death. This study initially selected the 23,598 longitudinal working-age (16 to 64 years-old) respondents from panel 2 who were in scope and responded to the labor interview in the first reference year. The first reference year for panel 2 was 1996 and the last reference year was 2001. For each reference year, information on labor is collected for respondents who are from 16 to 69 years of age on December 31 of the reference year.

---

[2] For further information on the SLID design and data processing refer to the SLID Microdata User's Guide (Statistics Canada, 1997).

**Table 1:** Annual percentages of attrition and non-response for the selected sample

| Wave | Attritors (out of scope) | | Non-Respondents to the labor interview (among in scope) | | Total | |
|------|-------------------|-------------------|-------------------|-------------------|---------|---------|
|      | Number (1) | Sample percentage (2) | Number (3) | Sample percentage (4) | (1)+(3) | (2)+(4) |
| 1997 | 647   | 2.74  | 1,319 | 5.59  | 1,966 | 8.33  |
| 1998 | 1,001 | 4.24  | 1,589 | 6.73  | 2,590 | 10.98 |
| 1999 | 1,629 | 6.90  | 2,047 | 8.67  | 3,676 | 15.58 |
| 2000 | 2,451 | 10.39 | 2,780 | 11.78 | 5,231 | 22.17 |
| 2001 | 2,949 | 12.50 | 2,211 | 9.37  | 5,160 | 21.87 |

Note: Includes the 23,598 longitudinal working-age (16 to 64 years-old) respondents who were in-scope and responded to the first wave of the labor interview in the 1996 reference year.

The 23,598 respondents initially selected for this analysis represent 17.7 million persons age 16 to 64 years old on December 31, 1996. The weight variable used for the analyses, except where noted otherwise, is the longitudinal person survey weight for the 1996 reference year (Levesque and Franklin, 2000). A longitudinal person survey weight for the 1996 reference year is assigned a value greater than zero for all persons selected in December, 1995 for the longitudinal sample who were in-scope and responded for the 1996 reference year. This longitudinal weight is adjusted for non-response and inter-provincial migration in the 1996 reference year. Longitudinal respondents who were out-of-scope at the end of the 1996 reference year or who were non-respondents have a survey weight value of zero and are excluded from our analysis. Moreover, the analysis here includes only longitudinal respondents who provided information in the labor interview to ensure that a minimum of information is available for comparing the characteristics of respondents who exit the sample in subsequent years to those of respondents who are still in scope and providing information at the end of the panel period.

The percentage of our study sample that became out of scope (attrition) reached 12.5% by the end of the panel (see Table 1). The other issue of concern in this analysis is complete non-response to the labor interview among respondents who were in-scope. Table 1 shows that complete non-response to the labor interview was at 9.37% of our study sample at the end of the panel. This percentage was the highest in 2000 (at 11.78% of our initial sample). In total, over one-fifth of the longitudinal sample who were in-scope and responded to the first wave of the labor interview were out of scope or non-respondents to the labor interview by the last reference year of the panel.

The sample used in the estimation of the model developed in Section 3, is of 22,990 longitudinal in-scope respondents who provided sufficient information in the labor interview and were age 16 to 64 years in 1996. By the end of 2001, 7,381 individuals or 32% of these respondents became attritors or non-respondents at least once. To ease the estimation of our model, we consider attrition as an absorbing state, so respondents, who became out of scope or did not respond to the labor force interview during a wave, are considered as attritors/non respondents for the subsequent waves. Only a minority, about one-fifth, of respondents who are out of scope or a non-respondent to the labor interview are converted to labor interview respondents in the following wave. This conversion dropped considerably in the latter half of the panel. In our sample, all cases were respondents to the labor interview for 1996; therefore conversions from out of scope and non-response begin for 1998.

The characteristics of attritors/non-respondents and respondents differ and suggest that attrition may not be random within the SLID longitudinal sample. Averages of the characteristics of the selected observations are presented in Table 2. The averages are calculated for in-scope labor-interview respondents for year (t) by the response status (respondent or attritor/non-respondent) for year (t+1).

**Table 2:** Characteristics of attritors/non-respondents and non-attritors
(Standard-deviations are reported in parentheses.)

| | | \multicolumn{5}{c}{Averages for year (t)} | | | | |
|---|---|---|---|---|---|---|---|---|---|
| | | 1996 | 1997 | 1998 | | 1999 | | 2000 | |
| Response status for year (t+1) | | | | | | | | | |
| Age** | Respondents | 38.3 | 39.4 | 40.6 | | 41.9 | | 42.6 | |
| | | (0.07) | (0.08) | (0.09) | | (0.09) | | (0.10) | |
| | Attritors | 37.2 | 36.6 | 37.5 | | 38.5 | | 40.0 | |
| | | (0.41) | (0.44) | (0.42) | | (0.35) | | (0.52) | |
| Female | Respondents | 0.504 | 0.505 | 0.508 | * | 0.511 | * | 0.508 | |
| | Attritors | 0.499 | 0.500 | 0.478 | * | 0.475 | * | 0.483 | |
| Married** | Respondents | 0.628 | 0.638 | 0.652 | | 0.671 | | 0.672 | |
| | Attritors | 0.526 | 0.473 | 0.519 | | 0.512 | | 0.528 | |
| Child at home** | Respondents | 0.179 | 0.172 | 0.164 | | 0.159 | | 0.149 | |
| | Attritors | 0.130 | 0.138 | 0.125 | | 0.131 | | 0.112 | |
| Urban area** | Respondents | 0.821 | 0.813 | 0.803 | | 0.802 | | 0.805 | |
| | Attritors | 0.875 | 0.887 | 0.884 | | 0.831 | | 0.842 | |
| Moved during year** | Respondents | 0.150 | 0.132 | 0.143 | | 0.131 | | 0.145 | |
| | Attritors | 0.189 | 0.220 | 0.196 | | 0.217 | | 0.228 | |
| Immigrant | Respondents | 0.182 * | 0.175 * | 0.171 | * | 0.170 | * | 0.171 | |
| | Attritors | 0.242 * | 0.285 * | 0.223 | * | 0.201 | * | 0.210 | |
| Student | Respondents | 0.200 | 0.182 * | 0.156 | * | 0.122 | * | 0.117 | * |
| | Attritors | 0.215 | 0.238 * | 0.218 | * | 0.178 | * | 0.150 | * |
| Employed during year** | Respondents | 0.770 * | 0.782 | 0.784 | | 0.778 | | 0.786 | * |
| | Attritors | 0.739 * | 0.752 | 0.763 | | 0.766 | | 0.740 | * |
| Weeks employed** | Respondents | 36.4 | 37.1 | 37.5 | | 37.8 | | 38.4 | |
| | | (0.19) | (0.20) | (0.21) | | (0.21) | | (0.21) | |
| | Attritors | 34.1 | 34.5 | 35.5 | | 35.5 | | 35.0 | |
| | | (0.76) | (0.84) | (0.81) | | (0.68) | | (0.88) | |
| Wages and Salaries ($)** | Respondents | 25,092 | 26,693 | 27,840 | | 29,641 | | 32,311 | |
| | | (159) | (188) | (216) | | (221) | | (256) | |
| | Attritors | 23,342 | 24,035 | 26,105 | | 26,280 | | 28,314 | |
| | | (642) | (867) | (759) | | (791) | | (1,337) | |
| Composite hourly wage ($)** | Respondents | 14.68 | 15.27 | 15.76 | | 16.63 | | 17.69 | |
| | | (0.08) | (0.08) | (0.09) | | (0.10) | | (0.11) | |
| | Attritors | 13.81 | 13.76 | 14.62 | | 14.79 | | 15.64 | |
| | | (0.25) | (0.35) | (0.32) | | (0.27) | | (0.47) | |
| Household income from all sources ($) | Respondents | 58,572 * | 60,628 * | 62,840 | | 65,799 | * | 69,317 | |
| | | (562) | (716) | (688) | | (922) | | (781) | |
| | Attritors | 54,172 * | 54,669 * | 61,805 | | 59,699 | * | 65,693 | |
| | | (1,669) | (2,074) | (3,406) | | (2,010) | | (4,592) | |

Note: All estimates are weighted using the longitudinal person weight for the reference year (t). In parentheses are standard-errors estimated using balanced repeated replication with the bootstrap replicate weights provided by Statistics Canada.
* The difference in the means between respondents and attritors is statistically significant at the 5 percent level.
** The difference in the means for all years is statistically significant at the 5 percent level.

For each year, people who become attritors or non-respondents in the subsequent year are on average younger, and are likely to be unmarried and immigrants compared to those who remain in the sample. In addition, respondents who are living in an urban area or who moved during the reference year (t) are more likely to become attritors in year (t+1). Moreover, attritors are more likely to live in urban areas with larger populations in their residential area than non-attritors. Attrition is also higher when respondents live without a spouse or common-law partner. It follows that on average, respondents who would later become attritors have lower wages and salaries and lower total household income. Despite these differences between the two groups, what actually matters here is not that attrition is not random but rather it may not be random even after controlling for the observable characteristics.

## 3. ECONOMETRIC SPECIFICATION

Consider a longitudinal survey with $T$ waves. At time period $t = 1$, a probabilistic sample is obtained from the target population. At time $t \geq 2$, some individuals are ineligible (i.e., out of scope) for the sample. In this study, we refer to these individuals as "attritors." Individuals who can not be located or contacted for an interview and those persons who refuse to respond to the labor questionnaire when contacted are referred to here as "non-respondents". For convenience, we will be using attrition to indicate both attrition and non response in the rest of the text. The model focuses on the possible correlation between three variables: wages, employment status and respondent status. Wages are only observed for employed respondents, and the employment status is only observed for respondents who provide information on their employment during the reference year. Therefore, data on employment status is censored (missing) for attritors, and wages are censored for attritors and for respondents who are not employed. If this two-level censorship is not random after controlling for observables, results based on observed data are subject to selection bias. In order to evaluate this potential selection bias when estimating employment status and wages, we propose the following model. The two selection sources are depicted by the reduced-form equations (1) and (2) below:

Non-attrition criterion:

$$a_{it}^* = Z_{i(t-1)}\theta + \varepsilon_{1it}, \; i=1,...,n; \; t=2,...,T_i \qquad (1)$$

where $i$ indexes for individuals and $t$ indexes for time periods (i.e. waves of the survey). Individual $i$ is a respondent in period $t$ ($a_{it} = 1$) if $a_{it}^* \geq 0$, and is an attritor in period $t$ ($a_{it} = 0$) otherwise. Because of the assumption that attrition is an absorbing state, if $a_{it} = 0$, then $a_{it'} = 0$ for any $t'>t$. Since information for the current period is not available for attritors, we use lagged variables in Equation (1). The initial period of analysis of attrition is the second wave since all individuals respond in the first wave.

Employment criterion:

$$e_{it}^* = X_{it}\alpha + \varepsilon_{2it}, \; i=1,...,n, \; t=1,...,T_i \qquad (2)$$

Respondent $i$ is employed ($e_{it} = 1$) if $e_{it}^* \geq 0$, and is not employed ($e_{it} = 0$) if $e_{it}^* < 0$. A person is not employed if she is either unemployed or not in the labor force. $Z_{it}$ and $X_{it}$ are vectors of covariates, and $\varepsilon_{1it}$ and $\varepsilon_{2it}$ are random components capturing unobserved variables. $Z_{it}$ and $X_{it}$ are observed for all individuals in the sample in the first time period (i.e., for $t = 1$), and then are observed whenever $a_{it} = 1$. Similar to Zabel (1998), we include wave dummies in $Z_{it}$ and $X_{it}$ to account for duration dependence. A monotonic change in the coefficients on the wave dummies indicates the presence of such dependence. In Equation (1), negative dependence suggests that the probability of attrition from the survey is increasing over time, *ceteris paribus*. In other words, the likelihood of an individual being observed in the sample decreases over time. On the other hand, positive duration implies that survey participants likely remain as in-scope respondents for the duration of the panel's reference period.

Wages equation:

Wages are given by the following equation:

$$y_{it} = W_{it}\beta + \varepsilon_{3it}, \quad i = 1,\ldots, n, \, t = 1,\ldots, T_i \quad (3)$$

where $y_{it}$ is log hourly wage,[3] $W_{it}$ is a vector of exogenous covariates, and $\varepsilon_{3it}$ is a random component. The structural model is given by Equations (1), (2) and (3). This model is sequential since dummy variable $e_{it}$ is observed only if $a_{it} = 1$ (the individual responds), and $y_{it}$ is observed only if $a_{it} = 1$ and $e_{it} = 1$ (the individual responds and is employed[4]) (see Maddala, 1983, pp. 278-283, for further examples on multiple criteria for selectivity).

Ideally, one would like to estimate the model by considering random terms $\varepsilon_{jit}$, $j=1,2,3$, $t = 1,\ldots,T_i$, to be freely correlated for the same individual. However, doing so will involve computing joint probabilities from a $3xT_i$ variate distribution, which is practically problematic. In order to ease the estimation of the model, we will adopt the random effect model approach (see below). We also estimate the model consistently in two stages following the approach suggested by Ham (1982). The latter approach is an extension of the two-stage estimator for the one selection rule proposed by Heckman (1979), and is computationally more attractive than the maximum likelihood method. The first stage involves a joint estimation of the selection equations (1) and (2) using panel data.[5] Then, correction terms using obtained parameter estimates are calculated and inserted in the wages equation (3) to account for selection bias.

---

[3] In the empirical estimation, we consider the composite hourly wage for all paid-worker jobs held by the respondent during year $t$.
[4] Notice that we estimate Equation (3) using hourly wages, which are given only for paid workers. However, employed workers include non-paid workers. So, the latter are ignored in Equation (3).

Stage 1: Estimation of selection equations

In order to simplify the computation of joint probabilities, we adopt the random effects model, which specifies:

$$\varepsilon_{1it} = u_{1i} + v_{1it} \qquad \varepsilon_{2it} = u_{2i} + v_{2it} \qquad \varepsilon_{3it} = u_{3i} + v_{3it} \qquad (4)$$

where $u_{1i}$, $u_{2i}$ and $u_{3i}$ are individual specific effects assumed to be freely correlated, but independent of $Z_{it}$, $X_{it}$ and $W_{it}$, and of $v_{jit}$ for $j=1,2,3$ and $t = 1,...,T_i$. We also assume that error terms $v_{jit}$ are independently distributed over individuals and time. In addition, $v_{jit}$ are mutually independent. Hence, the correlations between Equations (1), (2) and (3) are given by the correlations between individual specific effects $u_{1i}$, $u_{2i}$ and $u_{3i}$. Let $u_i^* = (u_{1i}, u_{2i}, u_{3i})'$. Conditional on $u_i^*$, $\varepsilon_{1it}$, $\varepsilon_{2it}$ and $\varepsilon_{3it}$ are independent. The vector $u_i^*$ is assumed to follow a trivariate normal distribution:

$$u_i^* \sim N(0, \Sigma) \qquad (5)$$

where $\Sigma = \begin{pmatrix} \sigma_{11} & \sigma_{12} & \sigma_{13} \\ & \sigma_{22} & \sigma_{23} \\ & & \sigma_{33} \end{pmatrix}$.

Attrition is random and there is no selectivity bias in employment equation estimates (Equation 2) if unobserved individual determinants of employment are uncorrelated with unobserved determinants of attrition (i.e. if $\sigma_{12} = 0$). Likewise, there is no selectivity from attrition when estimating the wage equation if $u_{1i}$ and $u_{3i}$ are uncorrelated (i.e. if $\sigma_{13} = 0$). As described

---

[5] Ham (1982) uses only cross-sectional data.

above, the first stage of our procedure involves the joint estimation of Equations (1) and (2). For this purpose, the individual contribution to the likelihood function conditional on $u_i = (u_{1i}, u_{2i})'$

is: $L_i(u_i) = \prod_{t=1}^{T_i} L_{it}(u_i)$ (6)

where

$$L_{it}(u_i) = \{Pr(a_{it} = 0 | u_i)\}^{1-a_{it}} \times \left[\{Pr(a_{it} = 1, e_{it} = 0 | u_i)\}^{(1-e_{it})} \times \{Pr(a_{it} = 1, e_{it} = 1 | u_i)\}^{e_{it}}\right]^{a_{it}}$$ (6.1)

for $t \geq 2$, and $L_{it}(u_i) = L_{i1} = \{Pr(e_{it} = 0 | u_i)\}^{(1-e_{it})} \times \{Pr(e_{it} = 1 | u_i)\}^{e_{it}}$ for $t=1$. (6.2).

Since all individuals responded to the labor interview at time period $t=1$, the contribution of an individual to the likelihood function depends only on the employment status at this period. Given that $\varepsilon_{1it}$ and $\varepsilon_{2it}$ are independent conditional on $u_i$, Equation (6.1) simplifies to:

$$L_{it}(u_i) = \{Pr(a_{it} = 0 | u_i)\}^{1-a_{it}} \times \left[Pr(a_{it} = 1 | u_i)(Pr(e_{it} = 0 | u_i))^{1-e_{it}} (Pr(e_{it} = 1 | u_i))^{e_{it}}\right]^{a_{it}}$$ (6.3)

For identification purposes, we shall assume that $v_{1it}$ and $v_{2it}$ are $N(0,1)$ distributed. Thereafter, the unconditional contribution of an individual to the likelihood function is:

$L_i = \int_{-\infty}^{\infty} \int_{-\infty}^{\infty} L_i(u_i) g(u_i) du_{1i} du_{2i}$ (7)

where $g(.)$ is the joint density function of $u_{1i}$ and $u_{2i}$. Finally, full maximum likelihood estimates of the parameters in (1) and (2) are obtained by maximizing the log likelihood

function: $log(L) = \sum_{i=1}^{n} log(L_i)$ (8)

Since the function in (8) involves two-dimensional integration, direct optimization is generally not feasible. We rather use maximum simulated likelihood. Notice that the function in (7) is an expectation ($L_i = E_{u_i}[L_i(u_i)]$), which can be approximated by a simulated mean:

$$L_{is} = \frac{1}{R}\sum_{r=1}^{R} L_i(u_{ir}) \qquad (9)$$

where $u_{ir}$, $r = 1,\ldots,R$, are $R$ draws from the bivariate distribution of $u_i$. $u_{1i}$ and $u_{2i}$ can be specified as linear combinations of two independent $N(0,1)$, $\eta_{1i}$ and $\eta_{2i}$:

$$u_{1i} = s_1\eta_{1i} + s_2\eta_{2i} \text{ and } u_{2i} = s_3\eta_{2i} \qquad (10)$$

$s_1$, $s_2$ and $s_3$ are three unknown coefficients to be estimated. Notice that $u_{1i}$ and $u_{2i}$ are independent if $s_2 = 0$. Finally, parameters in (1) and (2) including $s_1$, $s_2$ and $s_3$ are obtained by maximizing the simulated log likelihood:[6] $\log(L_s) = \sum_{i=1}^{n} \log(L_{is}) \qquad (11)$

A sample from $u_i$ is constructed as follows. First, we draw two independent samples of size R each from a $N(0,1)$. Then, a sample from $u_i$ is obtained using formulas in (10). Gourieroux and Monfort (1996) show that if $\sqrt{n}/R \to 0$ and $R$ and $n \to \infty$, then the maximum simulated likelihood estimator and the true maximum likelihood estimator are asymptotically equivalent. In the empirical application, we use $R = 50$.[7]

Stage 2: Estimation of selection-adjusted wages equation

The expectation of $y_{it}$ conditional on responding and being employed (and ignoring correlation across observations) is: $\quad E(y_{it} \mid a_{it}^* \geq 0, e_{it}^* \geq 0) = W_{it}\beta - \sigma_{13}\lambda_{1t} - \sigma_{23}\lambda_{2t} \qquad (12)$

where

$$\lambda_{1t} = \phi\left(Z_{i(t-1)}\frac{\theta}{\sqrt{1+\sigma_{11}}}\right)\Phi(Z_{it}^*)/P_{it}; \quad \lambda_{2t} = \phi\left(X_{it}\frac{\alpha}{\sqrt{1+\sigma_{22}}}\right)\Phi(X_{it}^*)/P_{it}$$

---

[6] See Gourieroux and Monfort (1996) and Train (2002) for discussion and statistical background. See also Green (2002) for some applications of the maximum simulated likelihood.
[7] We initially estimated the model using $R = 30$. There is little change in the results when increasing the number of draws from 30 to 50. Nevertheless, given the complexity of the likelihood function and the large size of the sample, the estimation of the model is computationally demanding even with a small number of draws.

$$Z_{it}^* = \left( X_{it} \frac{\alpha}{\sqrt{1+\sigma_{22}}} - \rho Z_{i(t-1)} \frac{\theta}{\sqrt{1+\sigma_{11}}} \right) / \sqrt{1-\rho^2}$$

$$X_{it}^* = \left( Z_{i(t-1)} \frac{\theta}{\sqrt{1+\sigma_{11}}} - \rho X_{it} \frac{\alpha}{\sqrt{1+\sigma_{22}}} \right) / \sqrt{1-\rho^2}$$

$$\rho = corr\left( \frac{\varepsilon_{1it}}{\sqrt{1+\sigma_{11}}}, \frac{\varepsilon_{2it}}{\sqrt{1+\sigma_{22}}} \right) = \frac{\sigma_{12}}{\sqrt{(1+\sigma_{11})(1+\sigma_{22})}}$$

$$P_{it} = F\left( Z_{i(t-1)} \frac{\theta}{\sqrt{1+\sigma_{11}}}, X_{it} \frac{\alpha}{\sqrt{1+\sigma_{22}}}, \rho \right) = \Pr(a_{it}=1, e_{it}=1)$$

for $t \geq 2$ (see Ham, 1982). For time period $t = 1$, there is only one source of selection which is employment status. For this period, $\lambda_{2t}$ is simply the inverse Mills ratio and $\lambda_{1t}$ is set to 0. Parameter estimates from the first stage are used to form consistent estimates $\hat{\lambda}_{1t}$ and $\hat{\lambda}_{2t}$ of $\lambda_{1t}$ and $\lambda_{2t}$. Then we estimate $\beta$, $\sigma_{13}$ and $\sigma_{23}$ by running the pooled OLS regression using the selected sample (as suggested by Wooldrige, 2001, Chapter 17, for a model with one selection criterion):

$$y_{it} = Z_{it}\beta - \sigma_{13}\hat{\lambda}_{1it} - \sigma_{23}\hat{\lambda}_{2it} + \varepsilon_{3it}^* \qquad (13)$$

where $\varepsilon_{3it}^* = \varepsilon_{3it} + \sigma_{13}\left(\hat{\lambda}_{1it} - \lambda_{1it}\right) + \sigma_{23}\left(\hat{\lambda}_{2it} - \lambda_{2it}\right)$.

Consistent estimates of the standard errors of the OLS slopes are obtained using formulas from Ham (1982).

As a final point, a question of identification of wage equation coefficients naturally arises. This identification requires exclusion restrictions, i.e. instruments that can predict attrition and employment decisions without directly affecting the wages. In the estimated model presented in Section 4, several variables in the selection equations are excluded from the wage equation. For instance, family size, ownership of the dwelling, and geographical mobility are assumed to affect attrition but not wages. Also, being a student, having a preschool child (for women), and non-labor income are assumed to affect employment but not wages. Yet, even without instruments, wage equation might still be identified off the nonlinearity - in Eq. 12, $\lambda_{1t}$ and $\lambda_{2t}$ are nonlinear functions of $Z_{it}$ and $X_{it}$ (Willis and Rosen, 1979).

## 4. EMPIRICAL RESULTS

Structural parameters estimates are obtained following the procedure described in Section 3 and are presented in Tables 3 and 4.

### 4.1 Non-Attrition and Employment Equations

At the outset, we notice that the estimated coefficient $s_2$ on $\eta_{2i}$ in Equation (10) is statistically significant at the 5 percent level, which means that the unobserved individual determinants of non-attrition and employment, $u_{1i}$ and $u_{2i}$, are correlated. The correlation between these terms is estimated at 0.45, but the estimated correlation between the whole random terms, $\varepsilon_{1it}$ and $\varepsilon_{2it}$, is only 0.03, but it is statistically significant at the 5 percent level. Hence, estimating non-attrition and employment equations separately will introduce very limited biases.[8] The positive and significant correlation between being a labor interview respondent and being employed indicates that, conditional on observed characteristics, respondents are more likely to be employed than are attritors. Employed workers are naturally less mobile because of their work attachment and, consequently, are relatively easier to locate especially in the subsequent waves of the survey.

We also conclude from Table 3 that variables which increase work attachment and/or reduce mobility (for instance, education, being married, and health status) also increase the likelihood of remaining in scope and responding. Also, we find that being an immigrant, and especially being an immigrant member of a visible minority group, reduces significantly both the probability of being employed and the probability of remaining a respondent in the sample. Moreover, a person who moved during a year (a signal of geographical mobility) is more likely to become an attritor in the subsequent year. The study by Zabel (1998)[9] using U.S. data leads to comparable results, with attritors having lower labor force attachment compared to individuals who continue to be survey respondents. The study also finds the same direction of the effects of some variables on attrition especially with regard to the effects of education and moving during the previous year.

---
[8] We estimated non-attrition and employment equations separately and we obtained estimates very close to those when the equations are estimated jointly.

Our results further indicate that females are less likely to attrite, but are less likely to be employed compared to males. Furthermore, being a student does not affect survey participation, though it reduces employment. Interestingly, however, age, ownership of the dwelling and family size, have no significant effects on attrition. Intuitively, one would expect that increased age, the possession of the dwelling or living in a large family, will reduce mobility, which, in turn, would increase the survey participation. By province, residents of Ontario are the most likely to attrite, followed by residents of Quebec. Incidentally, the latter are the least likely to be employed.

Another fact that is made obvious by our results is that increased family income lowers the likelihood of attrition. This result is in agreement with MaCurdy, Mroz, and Gritz (1998) who find that individuals who exit from the National Longitudinal Survey of Youth (NLSY) come disproportionately from the low income population. Finally, coefficient estimates on wave dummies support neither positive duration dependence nor negative duration dependence, since these estimates do not change monotonically. The likelihood of attrition is the highest in 2000 and the lowest in 1998, a fact that agrees with descriptive statistics presented in Table 1.

---

[9] Zabel (1998) analyzes attrition behaviour in the Panel Study of Income (PSID) and the Survey of Income and Program Participation (SIPP) in the U.S.A.

**Table 3:** Estimated Censored Bivariate Selection Model

|  | Non-attrition Equation | | Employment Equation | |
|---|---|---|---|---|
|  | Estimates | Std. err. | Estimates | Std. err. |
| Constant | 0.8029* | 0.1473 | -1.1922* | 0.3678 |
| Age / 10 | -0.0021 | 0.0489 | 3.015* | 0.1407 |
| Age squared / 100 | 0.0068 | 0.0056 | -0.4465* | 0.0167 |
| Female | 0.0346** | 0.0159 | -1.007* | 0.0601 |
| Immigrant | -0.1059* | 0.0301 | -0.1436 | 0.1101 |
| Immigrant x visible minority group | -0.1285* | 0.0413 | -0.461* | 0.1512 |
| Student | 0.0221 | 0.0273 | -0.7122* | 0.0499 |
| Married | 0.1751* | 0.0273 | 0.4068* | 0.0717 |
| Single | -0.0442 | 0.0328 | -0.1853*** | 0.1044 |
| Female x preschool child | - | - | -1.1864* | 0.0729 |
| High school graduate | 0.043*** | 0.0226 | 0.8015* | 0.0598 |
| Non university graduate | 0.0903* | 0.0235 | 1.3324* | 0.0741 |
| University graduate | 0.0968* | 0.0292 | 1.6728* | 0.098 |
| Household size | 0.0098 | 0.0068 | - | - |
| Urban areas | 0.1724* | 0.052 | -0.297*** | 0.1749 |
| log urban population size | -0.0579* | 0.0098 | 0.0396 | 0.0338 |
| Dwelling owner | -0.0145* | 0.0038 | - | - |
| Moved during the reference year | -0.1658* | 0.0224 | - | - |
| log household income | 0.0805* | 0.0224 | - | - |
| log non-labor income | - | - | -0.3469* | 0.0335 |
| Local unemployment rate | - | - | -0.0638* | 0.0084 |
| Health status[10] | 0.0295* | 0.0084 | 0.2341* | 0.0154 |
| Province of residence: | | | | |
|    Quebec | -0.0694* | 0.0242 | -0.2043** | 0.0799 |
|    Ontario | -0.1434* | 0.0219 | 0.0931 | 0.0797 |
|    Prairies | -0.0188 | 0.0239 | 0.4421* | 0.1008 |
|    BC | -0.0182 | 0.0301 | 0.1529 | 0.1003 |
| Wave dummies: | | | | |
|    1997 | - | - | 0.0804* | 0.0289 |
|    1998 | 0.1633* | 0.0245 | 0.0292 | 0.0355 |
|    1999 | 0.0461** | 0.0241 | -0.1227* | 0.0408 |
|    2000 | -0.1607* | 0.0233 | -0.0906*** | 0.0465 |
|    2001 | 0.1987* | 0.0269 | 0.0610 | 0.0473 |
| Parameters for the random terms (see Equation 10) | | | | |
|    $s_1$ | -0.0025 | 0.0373 | | |
|    $s_2$ | 0.0286** | 0.0124 | | |
|    $s_3$ | 2.2024* | 0.0445 | | |
| N | | | 22,990 | |

---

[10] This variable is coded as follows: 1=Poor, 2=Fair, 3=Good, 4=Very Good, 5=Excellent.

(*), (**) and (***): significant at the 1, 5 and 10 percent level (Two-tailed test).

With regard to employment determinants, it is important to note that the non-employed population includes unemployed as well as persons who are not in the labor force. Hence, a variable that decreases the probability of employment does not necessarily increases the probability of unemployment. Increased age, being male, increased schooling and living in the Prairie provinces, each increases the probability of employment. On the other hand the likelihood of being employed decreases for women, non-married persons, students and immigrants. Similarly, living in urban areas and regions with a higher unemployment rate; receiving increased levels of non-labor income also have negative effects on the probability of employment. Once again, coefficient estimates on wave dummies do not show a monotonic trend in the probability of being employed. The latter is the lowest in 1999 and the highest in 1997.

Finally, results suggest that unobserved heterogeneity plays a more important role in employment decisions than in attrition decisions. The estimated standard deviation of the individual random effect term is much larger in the employment equation (2.20 versus 0.03 for the non-attrition equation).

**4.2 Wage Equations**

Selection adjusted wage equation estimates are presented in Table 4. The dependent variable here is the composite hourly wage for all paid worker jobs during the reference years (1996 to 2001). SLID collects wage information for each paid worker job during the labor interview for respondents 16 to 69 years of age. The composite hourly wage is weighted by the total paid hours worked for each paid worker job. Potential years of experience are calculated as the difference between age at the time of the survey and estimated number of years of schooling corresponding to the highest education level of the respondent. We eliminate workers aged 16 with a high school diploma or more, aged 17 with more than high school, aged 18 to 21 with a Bachelors' degree or more, and aged 22 with a post-graduate university degree.

**Table 4:** Estimated selection adjusted wages equation

|  | Coef, | Std. Err. |
|---|---|---|
| Constant | 1.9446* | 0.0333 |
| Experience | 0.0308* | 0.0008 |
| Experience squared | -0.0004* | 0.0000 |
| Female | -0.1741* | 0.0048 |
| Immigrant | -0.0225** | 0.0090 |
| Immigrant x visible minority group | -0.0984* | 0.0145 |
| Health status[11] | 0.0158** | 0.0060 |
| High school graduate | 0.1070* | 0.0074 |
| Non university graduate | 0.1999* | 0.0095 |
| University graduate | 0.4935* | 0.0041 |
| Urban areas | -0.1861* | 0.0057 |
| log urban population size | 0.0231 | 0.0241 |
| Province of residence: |  |  |
|    Quebec | 0.1479* | 0.0062 |
|    Ontario | 0.2187* | 0.0058 |
|    Prairies | 0.0880* | 0.0071 |
|    BC | 0.2259* | 0.0057 |
| Wave dummies: |  |  |
|    1997 | 0.1734* | 0.0109 |
|    1998 | 0.1924* | 0.0071 |
|    1999 | 0.1970* | 0.0066 |
|    2000 | 0.2620* | 0.0125 |
|    2001 | 0.3354* | 0.0022 |
| Correction term from attrition | -1.0115* | 0.0388 |
| Correction term from employment | -0.2167* | 0.0195 |
| Adj. R-squared | 0.3897 |  |
| N:   - 1996 | 15 133 |  |
|    - 1997 | 14 440 |  |
|    - 1998 | 13 699 |  |
|    - 1999 | 12 454 |  |
|    - 2000 | 11 039 |  |
|    - 2001 | 10 525 |  |

Notes: the dependent variable is the log hourly wage. (*), (**) and (***): significant at the 1, 5 and 10 percent level (Two-tailed test).

---

[11] This variable is coded as follows: 1=Poor, 2=Fair, 3=Good, 4=Very Good, 5=Excellent.

In interpreting results, we focus on analyzing the biases that arise from ignoring non-attrition and employment selections on the estimation of the wage equation rather than analyzing the effects of covariates on the wage level. We mention that even when the correlation between the non-attrition and employment equations is set to zero, there is almost no effect on the adjusted wage equation estimates, a fact that confirms the mild dependence between the two statuses. The most novel result is that the coefficient estimates on the correction terms are negative and highly significant, indicating the non-randomness of both non-attrition and employment behaviors. From Equation (13), we can interpret the negative signs of these estimates as indications that wages are positively correlated with non-attrition and employment. However, the extent of selection bias from employment is larger than the extent of selection bias from attrition. The earnings gap between the selected (available) sample and a sample drawn randomly with identical observed characteristics is estimated at 9.65% due to non-attrition selection, and 13.51% due to employment selection.[12]

## 5. CONCLUDING REMARKS

The SLID and other longitudinal panel surveys are increasingly used by a diverse range of policy and academic researchers to investigate education, family, income and labor dynamics in Canada. Given the substantial impact that analyses using longitudinal survey microdata has on policy and academic knowledge, it is important for researchers to evaluate potential selection bias within their estimation. Statistics Canada continually assesses data quality and develops improved methodologies including survey weights to improve the reliability of population estimates.

---

[12] The gap is calculated by multiplying minus the selection coefficient times the mean value of the correction term.

This paper demonstrates that even when survey weights are used in model estimation, there is analytic value from assessing the potential selectivity from attrition and non-response. In this paper we specifically analyze the effects of attrition and non-response on the estimates from employment and wage equations. Our estimates provide evidence for the non-randomness of the attrition behavior. The correlation coefficient between random components in non-attrition and employment equations is positive and significant, though small. Generally speaking, the factors that increase the probability of being employed or less geographically mobile also increase the probability of being in scope for the survey and responding to the labor interview. These factors include increased education, being married, and being a non-immigrant. Furthermore, women are less likely to attrite compared to men. Finally, we find no indication of positive or negative duration dependence for responding to the survey, since the probability of responding to the labor interview does not evolve monotonically over time.

The wage equation estimates indicate significant selection biases from both non-attrition and employment. We also conclude that wages in the available longitudinal sample likely are higher than wages that would be observed if all respondents initially selected remained in the sample until the end of the panel. Similarly, we find that increased family income lowers the likelihood of attrition, which could result in a further over-estimation for family income. Hence, when using complex-design survey microdata, researchers can improve model estimation by evaluating the potential selectivity bias due to attrition and non-response within their models.